\documentclass[twocolumn]{emulateapj}
\usepackage{CJK}
\newcommand{\Msun}{$M_\odot$}
\newcommand{\Msunyr}{$M_\odot$yr$^{-1}$}

\newcommand{\uJy}{$\mu$Jy}
\newcommand{\uJyperbm}{$\mu$Jy\,beam$^{-1}$} 
\newcommand{\mstar}{$M_*$}

\newcommand{\kmpersec}{km\,s$^{-1}$}

\begin{document}
\begin{CJK*}{UTF8}{gbsn}



\title{{\it JWST} and ALMA imaging of dust-obscured, massive substructures in a typical $\lowercase{\it z} \sim 3$ star-forming disk galaxy}


\author{Wiphu Rujopakarn\altaffilmark{1,2},
Christina C. Williams\altaffilmark{3,4},
Emanuele Daddi\altaffilmark{5},
Malte Schramm\altaffilmark{6},
Fengwu Sun\altaffilmark{4},\\
Stacey Alberts\altaffilmark{4},
George H. Rieke\altaffilmark{4},
Qing-Hua Tan (谈清华)\altaffilmark{7},
Sandro Tacchella\altaffilmark{8,9},\\
Mauro Giavalisco\altaffilmark{10},
and John D. Silverman\altaffilmark{11}
}

\affil{$^1$National Astronomical Research Institute of Thailand, Don Kaeo, Mae Rim, Chiang Mai 50180, Thailand; wiphu@narit.or.th\\
$^2$Department of Physics, Faculty of Science, Chulalongkorn University, 254 Phayathai Road, Pathumwan, Bangkok 10330, Thailand\\
$^3$NSF's National Optical-Infrared Astronomy Research Laboratory, 950 North Cherry Avenue, Tucson, AZ 85719, USA\\
$^4$Steward Observatory, The University of Arizona, Tucson, AZ 85721, USA\\
$^5$Universit\'e Paris-Saclay, Universit\'e Paris Cit\'e, CEA, CNRS, AIM, 91191, Gif-sur-Yvette, France\\
$^6$Universit{\"a}t Potsdam, Karl-Liebknecht-Str. 24/25, D-14476 Potsdam, Germany\\
$^7$Purple Mountain Observatory, Chinese Academy of Sciences, 10 Yuanhua Road, Nanjing 210023, People's Republic of China\\
$^8$Kavli Institute for Cosmology, University of Cambridge, Madingley Road, Cambridge, CB3 0HA, UK\\
$^9$Cavendish Laboratory, University of Cambridge, 19 JJ Thomson Avenue, Cambridge, CB3 0HE, UK\\
$^{10}$Department of Astronomy, University of Massachusetts, B 619E 710 North Pleasant Street Amherst, MA, USA\\
$^{11}$Kavli IPMU (WPI), UTIAS, The University of Tokyo, Kashiwa, Chiba 277-8583, Japan\\
}




\begin{abstract}
We present an identification of dust-attenuated star-forming galactic-disk substructures in a typical star-forming galaxy (SFG), UDF2, at $z = 2.696$. To date, substructures containing significant buildup of stellar mass and actively forming stars have yet to be found in typical (i.e., main-sequence) SFGs at $z > 2$. This is due to the strong dust attenuation common in massive galaxies at the epoch and the scarcity of high-resolution, high-sensitivity extinction-independent imaging. To search for disk substructures, we subtracted the central stellar-mass disk from the {\it JWST}/NIRCam rest-frame 1.2 \micron\ image ($0\farcs13$ resolution) and subtracted, in the visibility plane, the central starburst disk from ALMA rest-frame 240 \micron\ observations ($0\farcs03$ resolution). The residual images revealed substructures at rest-frame 1.2 \micron\ co-located with those found at rest-frame 240 \micron, $\simeq 2$ kpc away from the galactic center. The largest substructure contains $\simeq20$\% of the total stellar mass and $\simeq5$\% of the total SFR of the galaxy. While UDF2 exhibits a kinematically-ordered velocity field of molecular gas consistent with a secularly evolving disk, more sensitive observations are required to characterize the nature and the origin of this substructure (spiral arms, minor merger, or other types of disk instabilities). UDF2 resides in an overdense region ($N \geqslant 4$ massive galaxies within 70 kpc projected distance at $z=2.690-2.697$) and the substructures may be associated with interaction-induced instabilities. Importantly, a statistical sample of such substructures identified with {\it JWST} and ALMA could play a key role in bridging the gap between the bulge-forming starburst and the rest of the galaxy.
\end{abstract}





\section{Introduction} \label{sec:intro}

The groundbreaking angular resolution and sensitivity of far-IR/submillimeter (submm) observations from the Atacama Large Millimeter/submillimeter Array (ALMA) and the subsequent flurry of blank-field interferometric submm surveys have enabled capturing an unbiased sample of {\it typical}\footnote{Typical star-forming galaxies are defined here as those with SFR within a factor of four of the main-sequence \citep[e.g.,][]{Rodighiero11} at their corresponding redshift on all of the following main-sequence parameterizations: \citet{Whitaker12, Speagle14}, and \citet{Schreiber15}, hereafter `typical SFGs'.} star-forming galaxies (SFGs) at $z \sim 1-4$ that represents the assembly sites of most stellar mass in the Universe independent of their dust attenuation \citep[e.g.,][]{Dunlop17, Elbaz18}. 

Sub-arcsecond ALMA observations of typical SFGs at this epoch have revealed a common pattern: they contain a compact region of intense star formation (SF) at their centers. These central star-forming regions usually harbor star-formation rate (SFR) of $\sim100-300$ \Msunyr\ within $r_{\rm SF} \sim 0.5-1$ kpc, which is $\sim 2-5$ times more compact than the extent of their stellar buildups virtually independently of the way they are selected, e.g., far-IR/submm blank field surveys \citep{Dunlop17,Fujimoto18,Elbaz18}, H$\alpha$-selected galaxies \citep{Tadaki17}, or optically-selected massive compact SFGs \citep{Barro16}.

\begin{figure*}[ht]
\figurenum{1}
\centerline{\includegraphics[width=\textwidth]{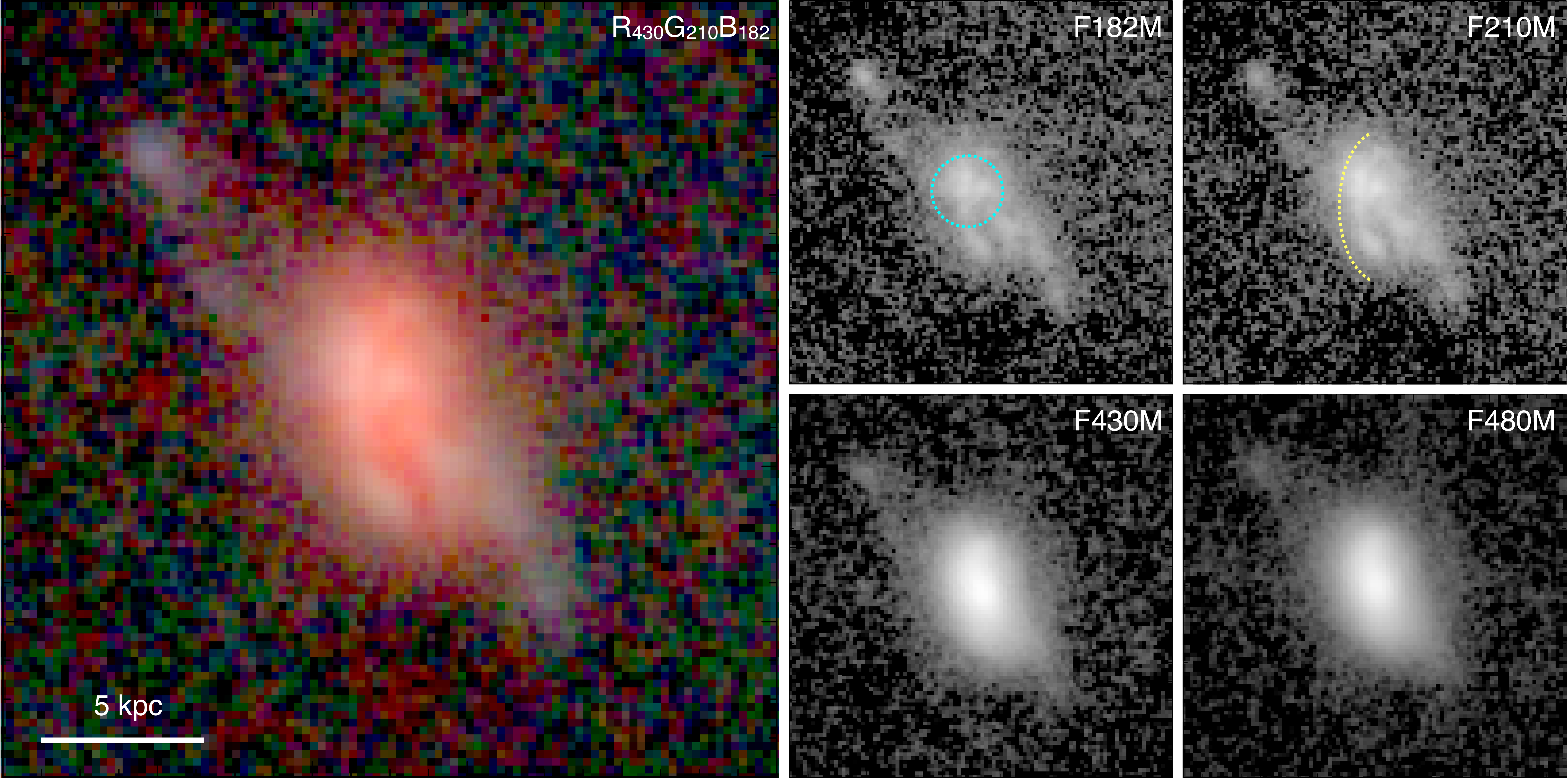}}
\caption{{\it JWST} images of UDF2. The left panel shows {\it JWST} color composite; the four sub-panels on the right show F182M image with the cyan dotted circle indicating the position of the brightest feature previously seen in {\it HST}/F160W image \citep[Figure 1 of][]{Rujopakarn19} that is now resolved into multiple point-like features; F210M image with the yellow dotted arc indicating the semicircular feature around the center of stellar mass distribution that is clearly visible in the F430M and F480M images. Each cutout is $3''\times3''$; north is up, east is to the left.
\label{fig:JWST_images}}
\end{figure*}

Zooming in further, high-S/N sub-0\farcs1 ALMA observations that can confirm or rule out substructures of SF are still scarce, but early results indicate that these central star-forming regions are strikingly smooth \citep{Rujopakarn19,Ivison20}. There has been no identification of off-center structures containing significant SFR in typical SFGs thus far --- ALMA identification of SF substructures at $z > 2$ so far are in luminous submm-selected galaxies \citep[e.g.,][]{Swinbank15, Iono16, Tadaki18, Hodge19}. Also, the UV-bright clumps commonly found in typical SFGs at $z \gtrsim 1$ contain $\lesssim$1\% of the total SF and molecular gas. That is, the UV-bright clumps are not the signposts for the `tip of the iceberg’ beneath which the bulk of SF is occurring \citep{Cibinel17, Rujopakarn19}.


For typical SFGs at $z > 2$, the connection between their compact core observed at far-IR/submm and their extended optical morphologies remains unclear. Bridging this gap requires finding off-center substructures that are detected in {\it both} far-IR/submm and optical bands, which represent the sites of SF and stellar mass buildups, respectively. Identification of off-center substructures with substantial presence of stellar mass and harboring significant SFR could indicate, e.g., spiral arms or other types of disk instability. Notably, rest-frame optical images of typical SFGs often exhibit strong dust attenuation in the proximity of the far-IR/submm emission \citep{Nelson19, Rujopakarn16, Rujopakarn19}, which suggest that {\it JWST} observations will play a key role in bridging the gap between optical and far-IR/submm pictures.

In this Letter, we report the first identification of off-center massive substructures in a typical SFG at $z \sim 3$ that is detected in both {\it JWST}/NIRCam and ALMA observations. We adopt a $\Lambda$CDM cosmology with $\Omega_M = 0.3$, $\Omega_\Lambda = 0.7$, $H_{0} = 70~{\rm km\,s}^{-1}{\rm Mpc}^{-1}$, where $1''$ at $z = 2.7$ corresponds to 7.9 kpc, and the Salpeter IMF. 


\begin{figure*}[ht]
\figurenum{2}
\centerline{\includegraphics[width=\textwidth]{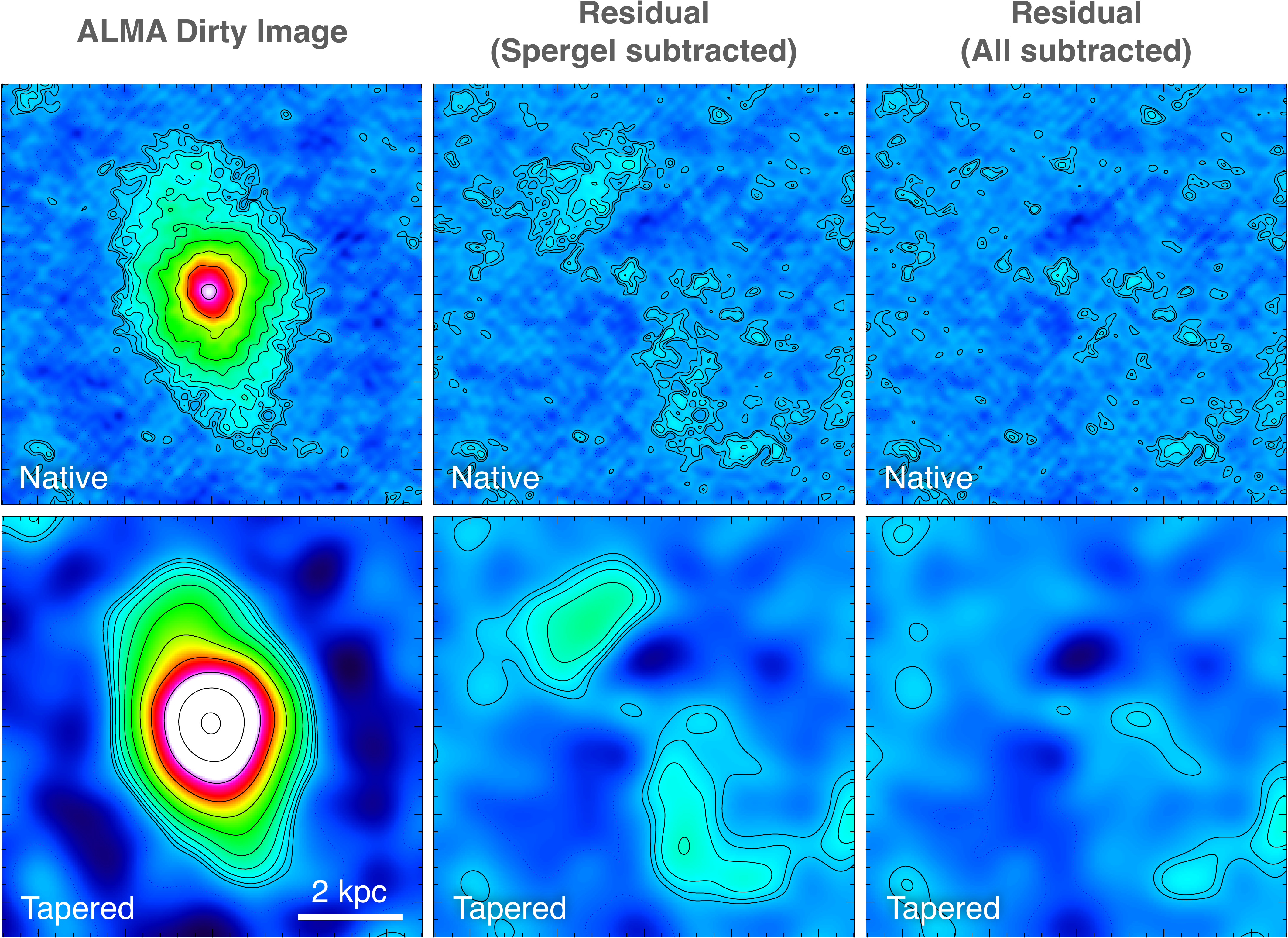}}
\caption{To search for substructures in UDF2's star-forming regions, we subtract from the ALMA rest-frame 240 \micron\ continuum visibilities a Spergel profile (an approximation of the Sersic profile with analytical Fourier transform). From left to right is the dirty ALMA image, the residual map with the main dust disk modeled by a single Spergel profile subtracted, and the residual after all components are subtracted, respectively. The top row is the native resolution image, whereas the bottom row is tapered at 1000 meters to enhance the sensitivity to extended features. Each cutout is $1''\times1''$; north is up, east is to the left; contours are $[-1.5, 1.5, 1.5^{1.5}, 1.5^{2}, ...] \times \sigma$.
\label{fig:ALMA_sub}}
\end{figure*}

\section{Target and Observations} \label{Target}

Our target (``UDF2'') was identified as a main-sequence SFG via an ALMA blank-field survey \citep{Dunlop17} of the Hubble Ultra Deep Field (HUDF). It has a total SFR of 260 \Msunyr\ and log(\mstar/\Msun) $= 10.9 \pm 0.2$ \citep{Rujopakarn19} at $z = 2.6961$ \citep{Kaasinen20}. {\it HST} imaging of UDF2 at $0.6-1.6$ \micron\ (rest-frame $0.2-0.4$ \micron) show disturbed morphology extending up to 2\farcs5 ($\simeq 20$ kpc) whereas the dust emission observed by ALMA originates from within the central 3 kpc region \citep{Rujopakarn16}. 

UDF2 is a subject of a sensitive search for SF substructure employing ALMA rest-frame 240 \micron\ dust continuum observations at $42 \times 30$ mas resolution and 9.6 \uJyperbm\ rms sensitivity \citep{Rujopakarn19}, which finds that the cold dust is distributed in a smooth disk with no substructures at the 200 pc scale. This data set remains one of the highest angular resolution imaging at the highest S/N of a typical SFG at $z > 2$ to date. 

\citet{Kaasinen20} further revealed by means of CO($3-2$) observations at 0\farcs88 $\times$ 0\farcs61 resolution that this disk is kinematically-ordered, well-described by a single rotating component with maximum rotational velocity, $V_{\rm rot,max} = 349^{+19}_{-23}$ \kmpersec\ and velocity dispersion, $\sigma = 77^{+10}_{-11}$ \kmpersec. The corresponding $V/\sigma$ ratio of $4.5 \pm 0.7$ is considerably more `dynamically cold' than disk galaxies at its epoch \citep[cf. compilation by][]{Rizzo20}. While a kinematically-ordered velocity field does not necessarily preclude mergers \citep[e.g.,][]{Simons19}, these observations lend credence to a picture of UDF2 being an isolated rotating disk, representative of typical SFGs at $z \sim 3$.

UDF2 was observed with {\it JWST}/NIRCam as a part of the JWST Extragalactic Medium-band Survey (JEMS; Program ID: 1963; PI: Williams). Observations were taken on 2022 October 11 in five filters, namely F182M, F210M, F430M, F460M, and F480M. The integration time was 28 ks in each filter (except F430M and F460M which were 14 ks each). The data reduction process is described in detail in \citet{Williams23}. The final mosaics are sensitive to 29.3, 29.2, 28.5, 28.3, and 28.6 AB mag (5$\sigma$, point source) in the five filters in the ascending order of wavelength and are resampled to 30 mas pixel size. We independently inspect the astrometry of the final mosaic with 30-mas ALMA images of the nearby UDF1, 3, and 7 \citep[whose astrometry is tied to the ICRF to $\simeq 2$ mas;][]{Rujopakarn19} and find that the offsets between brightest features in ALMA and {\it JWST} images are sub-pixel (i.e., $< 30$ mas).

\begin{figure*}[ht]
\figurenum{3}
\centerline{\includegraphics[width=\textwidth]{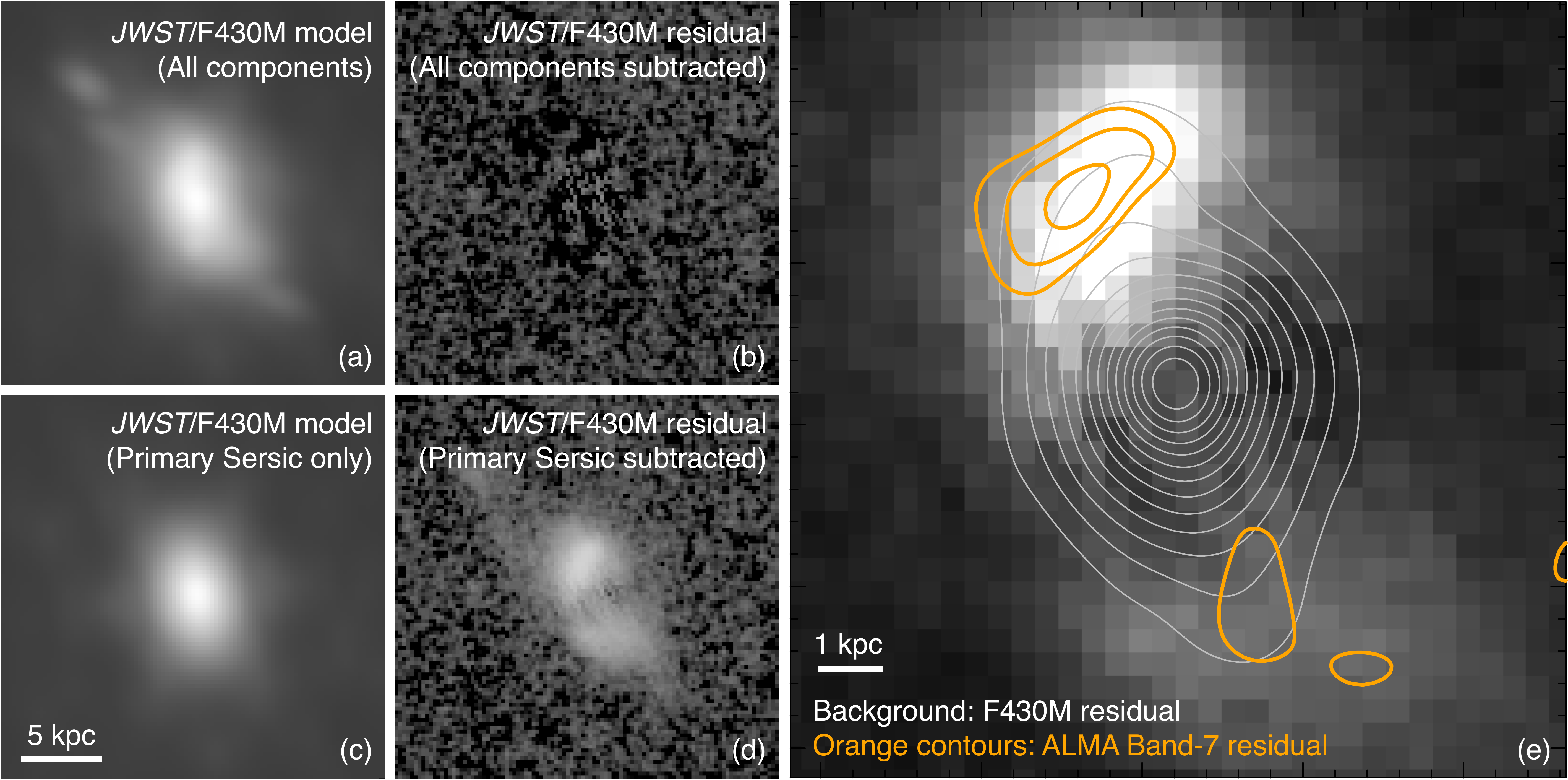}}
\caption{To search for substructures in the stellar mass distribution, we first model all components of the F430M emission (a), whose subtraction from the original image affords a flat residual (b), then only subtract the Primary S\'ersic model of the stellar mass disk (c). The residuals in (d) indicate the presence of stellar mass substructures northeast and southwest of the galaxy center. Cutouts in panels (a) to (d) are $3''\times3''$ and with a log scaling. Panel (e) shows the central $1''\times1''$ of the residuals in panel (d) with a linear scaling; the tapered ALMA residuals overplotted as orange contours starting from 3$\sigma$ in 1$\sigma$ increment. The grey contours show the pre-subtraction ALMA continuum, tapered to the same scale, starting from $3\sigma$ in 5$\sigma$ increment. The substructure in the northeast harboring SF and significant stellar mass buildup, dubbed the ``northern ridge'' is clearly visible.
\label{fig:JWST_sub}}
\end{figure*}


\section{Identification of Galactic-Disk Substructures} \label{Results}

To search for galactic-disk substructures, we model and subtract the central component from ALMA and {\it JWST} images to search for co-locating residuals in both bands. We note that the galaxy `center' refers to the centroid of the dust emission from the ALMA observations, whose position is localized within $\simeq$ 2 mas.
\begin{figure*}[ht]
\figurenum{4}
\centerline{\includegraphics[width=\textwidth]{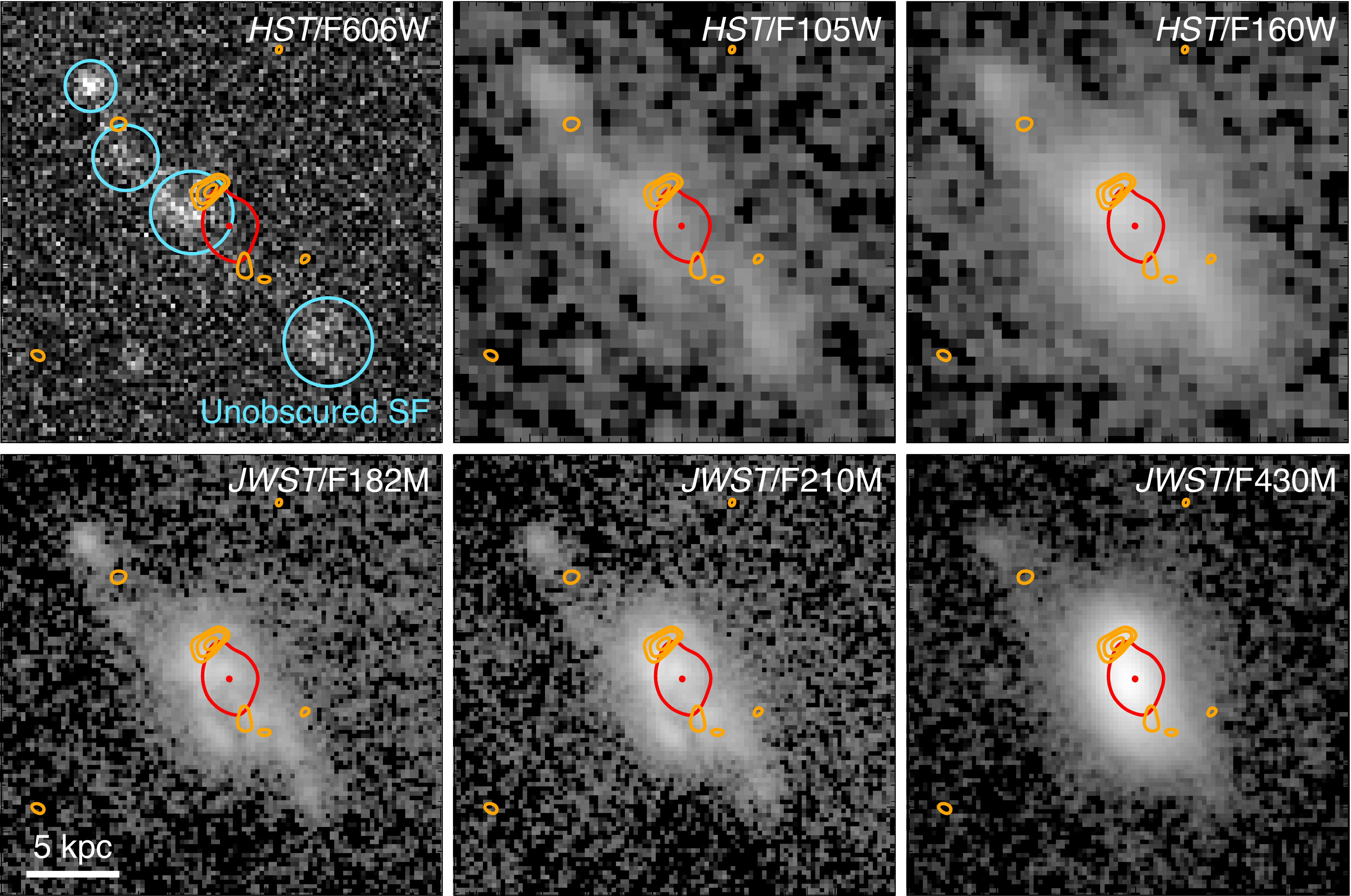}}
\caption{The star-forming substructures (orange contours starting from 3$\sigma$ in 1$\sigma$ increment), in the context of the rest of the galaxy. Notably, the {\it HST}/F606W image (rest-frame 160 \AA) shows the spatial dislocation between unobscured SF and the northern ridge; the {\it JWST}/F182M and F210M images show that the northern ridge appears to be a part of the semicircular arc around the central stellar mass distribution observed at F430M. The red contour delineates the 10$\sigma$ isophotal extent of unsubtracted ALMA emission (with the red dot marking the peak).
\label{fig:all_bands}}
\end{figure*}

\subsection{Rest-frame optical/near-IR morphology from JWST}

The {\it JWST} images of UDF2 are shown in Figure \ref{fig:JWST_images}. The most striking feature is the presence of a single stellar mass component, seen in all long-wavelength filters (F430M-F480M; rest-frame $1.2-1.3$ \micron), but not obvious at shorter wavelength bands. The long-wavelength images show an isolated disk with no major morphological disturbance. This represents the true stellar mass distribution, an observation that was not previously measurable in {\it HST}/F160W image because of the dust attenuation. The centroid and position angle of the stellar mass component is in close alignment with that of the cold dust disk and the rotation plane of the molecular gas disk \citep[Figure 7 of][]{Kaasinen20}. 

The F182M and F210M images (rest-frame $0.5-0.6$ \micron) show a more complex morphology including clumpy substructures surrounding the location of the stellar mass buildup. The F182M image contains a bright region 0\farcs2 northeast of the center, which corresponds to a prominent point-like feature in {\it HST}/F160W image (owing to the lower resolution). This bright region becomes less pronounced in F210M, where it appears to be a part of a semicircular feature (yellow arc in Figure 1) surrounding the location of the stellar mass buildup.

\subsection{Identification of star-forming substructures}\label{sec:ALMAsub}

To search for off-center cold dust substructures that could indicate galactic-disk SF, we re-analyze the ALMA rest-frame 240 \micron\ observations of UDF2 using the visibility-based modeling technique described in \citet{Rujopakarn19}. In the previous analysis, we modeled the SF disk using superpositioned Gaussians to represent the S\'ersic profile \citep{HoggLang13} because the latter lacks the analytical Fourier transform necessary for the visibility-based analysis. The best fit model was a double concentric Gaussian that affords an approximation of a S\'ersic. A region approximately 0\farcs2 northeast of the center contains some residual flux peaking at $\simeq2.9\sigma$, though the lack of definitively associable {\it HST} counterpart at its position \citep[see Figures 1 and 2 of][]{Rujopakarn19} precludes the recognition of its significance over other sub-$3\sigma$ noise peaks.

In this analysis, we employ the newly implemented \citet{Spergel10} profile in GILDAS version {\tt nov22a}, which offers an accurate representation of the S\'ersic profile without relying on multiple components and featuring an analytical Fourier transform. The Spergel profile is characterized by the Spergel index $\nu$, flux, and half-light radius \citep{Spergel10}, where Spergel $\nu$ plays a similar role to the S\'ersic $n$. Within a range of $\nu$, the Spergel profiles resemble S\'ersic profiles. For example, at $\nu=0.5$, the Spergel profile is identical to an exponential profile ($n=1$). The robustness of the GILDAS implementation of Spergel fitting was tested using simulations by \citet{Kalita22}; a robust mapping between the S\'ersic $n$ and Spergel $\nu$ will be presented in Q. Tan et al., in prep.

We fit all three galaxies in the ALMA beam (UDF1, 2, and 7) using the Spergel profile simultaneously using the GILDAS {\tt uv\_fit} task. The best-fit Spergel profile of UDF2 has major and minor axes of $135 \pm 3$ mas and $65 \pm 2$ mas, respectively, $\nu = -0.28 \pm 0.04$, and a flux of $2885 \pm 106$ \uJy. After subtracting the best-fit Spergel profile in the $uv$-plane, two regions of residual remain: one at the northeast and southwest of the galaxy center (Figure \ref{fig:ALMA_sub}). The northeast one, in particular, peaks at $5.5\sigma$ when Gaussian tapering is applied to baselines longer than 1000 meter (which results in a synthesized beam of $0\farcs16 \times 0\farcs13$ --- similar to that of {\it JWST}/F430M). We measure the fluxes of these substructures by fitting --- in the $uv$-plane --- elliptical Gaussians centering at the peaks of the substructure in the tapered image. The residual maps after subtracting the best-fit models, from which we have measured the substructure fluxes are shown in the `All subtracted' column in Figure \ref{fig:ALMA_sub}. The total flux in the northeast and southwest substructures are $172 \pm 28$ and $148 \pm 42$ \uJy, respectively, each being $\simeq$ 5\% of the total flux of the entire galaxy.

\subsection{Identification of stellar-mass substructures}\label{sec:JWSTsub}

To search for off-center stellar mass buildups that serve as a proxy for galactic disk substructures, we perform a GALFIT decomposition on the F430M image by modeling the source following the same procedure performed on the {\it HST} images in \citet{Rujopakarn19}. The F430M image is chosen for this analysis because it is the shortest band of the three long-wavelength module images, thereby affording the best spatial sampling of the substructures. The PSF model was derived from a median combination of non-saturating stars in the field.

The best-fit model (Figure \ref{fig:JWST_sub}) indicates a primary S\'ersic (containing 66\% of the total F430M emission) with $n = 0.98$, indicating that the central stellar mass buildup is well-described by an exponential disk; the half-light extent of the disk is $3.3\times2.3$ kpc with a position angle of 11$^{\circ}$ (counterclockwise from the north). Subtracting this exponential disk reveals two off-center substructures in the northeast and southwest of the center, which contain 23\% and 10\% of the total F430M flux, respectively; these three components altogether account for 99\% of the total model flux. The northeast substructure (which we will refer to as the `northern ridge' hereafter) is a part of (and has an elongated shape along) the semicircular feature observed at F210M.

Most significantly, the ALMA substructures are co-located with these F430M substructures, as shown in panel (e) of Figure \ref{fig:JWST_sub}, suggesting that they are off-center stellar mass substructures that harbor ongoing SF. Furthermore, the northern ridge appears to be partially overlapping with point-like features observed at F182M and F210M, which could represent the relatively less dust-attenuated (or patchy) emission from young stars (Figures \ref{fig:JWST_sub} and \ref{fig:all_bands}).

\subsection{Physical properties of the off-center substructures}

Having established the existence of off-center substructures containing both stellar mass and SF, we put constraints on their physical properties in order to aid the interpretation of their nature. We will focus on the northern ridge in this Letter and follow up with a more comprehensive study of UDF2 in the future when deeper {\it JWST}/NIRCam and MIRI imaging from the ongoing programs in Cycle 1 becomes available.

The northern ridge, shown in panel (e) of Figure 3, forms an arc spanning the position angle of $\simeq 0-50$ deg (counterclockwise from the north) of, and at a distance of $0\farcs20 - 0\farcs35$ ($1.6-2.8$ kpc) from the center. The length and width of the ridge are $\simeq 2.5 \times 1$ kpc. 

The ALMA rest-frame 240 \micron\ residual flux in the northern ridge is $\simeq 5\%$ of the total flux. If we assume that dust temperature and IMF do not vary spatially across the galaxy and that SFR varies linearly with the far-IR flux, the inferred SFR within the northern ridge is $\simeq 13$ \Msunyr. We note that the unobscured SFR inferred from rest-frame UV emission is $<1$ \Msunyr\ \citep{Dunlop17}; its contribution to the SFR of the northern ridge remains small, i.e., even the extended SF in the outer region of the galaxy is strongly dust-obscured. Likewise, the {\it JWST} rest-frame 1.2 \micron\ residual flux in the northern ridge is 23\% of the total flux. If we assume a constant IMF and that the stellar mass varies linearly with the 1.2 \micron\ flux, the stellar mass in the northern ridge is about log(\mstar/\Msun) of $10.2 \pm 0.2$. This stellar mass estimate should be taken as an upper limit considering the potential contribution from young stars to the rest-frame 1.2 \micron\ emission.

In summary, we have identified, within a typical SFG at $z = 2.7$, two off-center substructures containing $\simeq10-20$\% of total stellar mass and $\simeq$5\% of total SF. We highlight the fact that the central star-forming region is where $\simeq90\%$ of SF is located and also contains $\simeq$70\% of the total stellar mass, which implies that the substructures have $\sim2-4\times$ lower specific SFR (sSFR) compared to that of the center.


\section{Discussions} \label{Discussion}

There are notable parallels between the substructures identified in this work and those observed at $z \sim 1-4$ using {\it JWST}. For example, spiral arms have been identified optically by \citet{Wu23} using NIRCam observations of A2744-DSG-$z$3, a typical SFGs at $z = 3.06$, but the angular resolution and sensitivity of their ALMA observations are insufficient to resolve the spiral structures observed by {\it JWST}. Similarly, using NIRCam observations of seven sub-mm-selected typical SFGs at $z \sim 2$, \citet{Chen22} found spiral arms and bars and also nearby companions that suggest ongoing dynamical interactions (their IR/radio observations are $\sim$few arcsecond resolution, still not sufficient to associate the stellar mass and SF structures). \citet{Cheng23} observed 19 ALMA-selected cluster-lensed and field galaxies at $z \sim 1-4.5$ spanning both main-sequence and quiescent sSFR ranges using NIRCam and found substructures such as spiral arms to be common. Their ALMA observations were at $0\farcs3 - 2\farcs1$ angular resolution and start to reveal potential cases of SF within substructures and suggest that definitive associations of SF within off-center stellar-mass substructures could be commonplace with greater availability of high-resolution, high-S/N ALMA observations of NIRCam targets.

In both the \citet{Chen22} and \citet{Wu23} samples, companion and/or group environment appears to be a common attribute. We also find that UDF2 also resides in a potentially overdense region with four massive galaxies (including UDF2) identified within a projected distance of 70 kpc, spectroscopically confirmed (F. Sun, private communication) to be at $z = 2.690-2.697$ (Figure 5). These findings allude to a picture that the stellar-mass substructures with ongoing SF identified in UDF2 could represent a forming spiral arm or other types of disk instability resulting from minor merger or interactions in an overdense environment \citep[despite the overall cold gas velocity field appearing to be kinematically-ordered;][]{Kaasinen20}.

\begin{figure}
\figurenum{5}
\centerline{\includegraphics[width=\columnwidth]{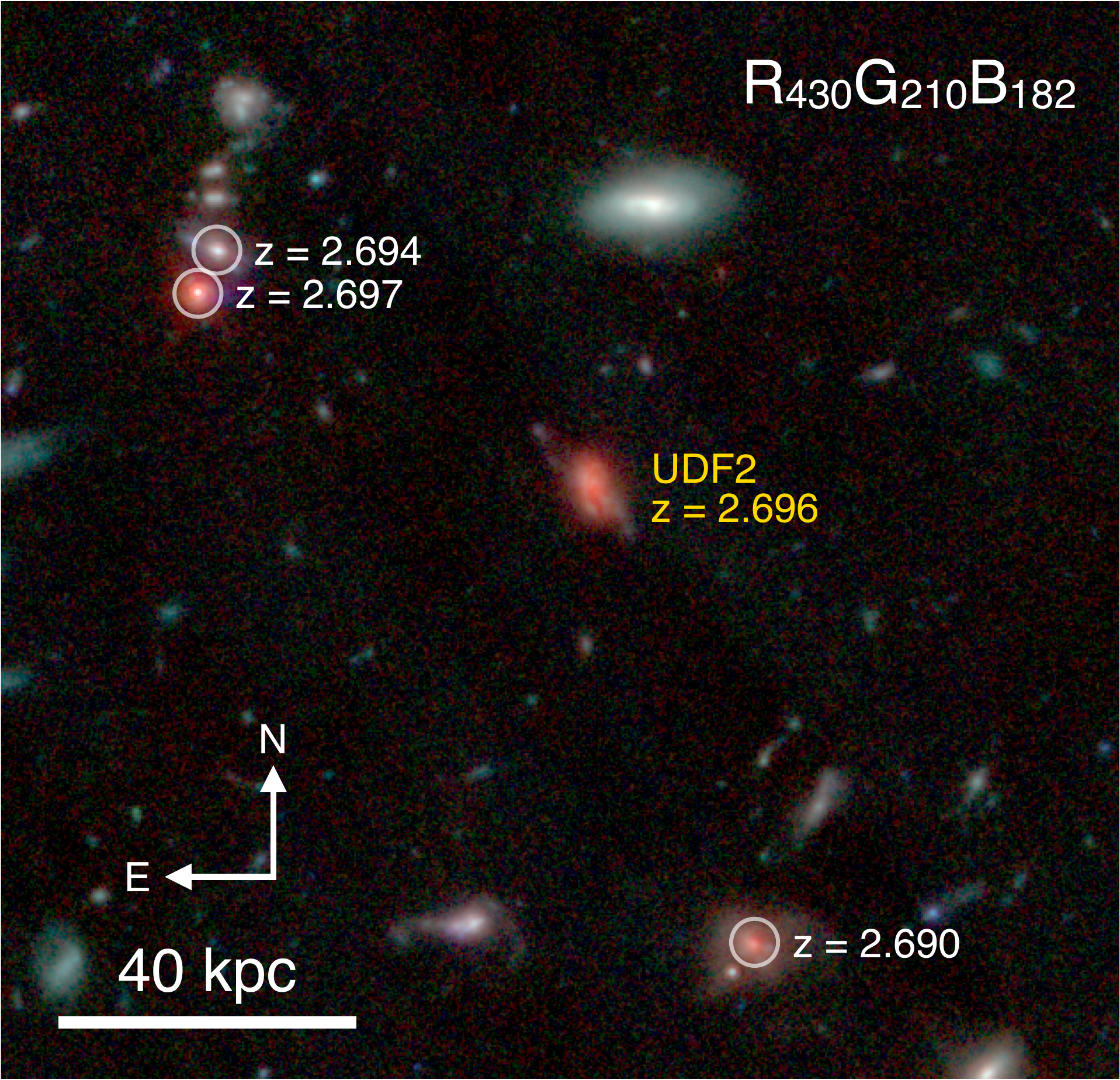}}
\caption{Three other massive galaxies spectroscopically confirmed to be at $z = 2.690-2.697$ are within a projected distance of 70 kpc from UDF2, suggesting a potentially overdense region that may have implications on the origin of the substructures identified.
\label{fig:environment}}
\end{figure}

On the contrary, local spiral arms contain stellar masses $\sim1-5\%$ of the disk \citep[e.g.,][]{Pichardo03}, significantly smaller than the substructures we have identified. The $2\times$ asymmetry of the stellar mass between the northern and southern ridges of UDF2 is another departure from classical spiral arms and bears a resemblance to the lopsided stellar disk reported by \citet{Kalita22}, who proposed that mass lopsidedness in the stellar disk could be a remnant indicating the `impact point' of the cold gas accretion stream (also in a cluster environment). Another similarity to the Kalita et al. substructure, which is quiescent, is that the northern ridge has $4\times$ lower sSFR compared to the core, and hence could represent an evolutionary stage prior to the Kalita et al. picture. The fueling and highly unstable disk \citep[e.g., a local analog of $z\sim1-2$ typical SFGs studied in detail by][]{Puschnig23} that has led to a structure like that of the northern ridge and the lopsided disks could be characteristic of typical SFGs at $z \sim 3$.

Simulations of gas-rich galactic disks have long predicted the existence of SF clumps, with the latest predictions being that they are $10^7-10^8$ \Msun\ in mass, SFR $\sim$1\% of galaxy-integrated SFR, and size $\sim50-100$ pc \citep[e.g.,][]{Mandelker17,Tamburello17,Faure21,Fensch21}. So far, the predictions are at odds with the handful of typical SFGs at $z \sim 2-3$ with high-resolution, high-S/N ALMA observations: the central SF regions of UDF2 and other HUDF galaxies are smooth and the strongly-lensed `Eyelash' is also clumpless down to 1\% SFR at $\simeq$80 pc resolution \citep{Rujopakarn19,Ivison20}. It is hence natural to ask if we are finally seeing off-center submm SF clumps here. Perhaps not directly: the northern ridge is significantly larger in stellar mass, SFR, and especially the size than the predicted clumps, though we cannot exclude a scenario that the northern ridge is a structure comprising multiple clumps of SF, which can be tested using higher resolution ALMA observations at higher S/N. 

This Letter highlights the importance of the combination of ALMA and {\it JWST} to probe dusty galactic structure at $z \sim 3$. Apparent voids at shorter wavelengths due to dust obscuration/attenuation are being filled by {\it JWST}. In the case of UDF2, the location of the bulk of the stellar mass buildup is revealed to be co-located with the starburst core, affording a clearer picture of the galactic bulge during its assembly, and also galactic-disk substructures actively forming stars. Building a statistical sample of disk substructures in typical SFGs with optical/submm counterparts will be crucial to help bridge the gaps between optical/near-IR and far-IR/submm morphologies, as well as between the compact core SF and the rest of the galaxy.

\section*{Acknowledgments}
We thank the anonymous referee for their helpful suggestions and comments. This Letter makes use of observations from {\it JWST} Cycle 1 GO program \#1963. Support for program JWST-GO-1963 was provided in part by NASA through a grant from the Space Telescope Science Institute, which is operated by the Associations of Universities for Research in Astronomy, Incorporated, under NASA contract NAS 5-26555. The research of CCW is supported by NOIRLab, which is managed by the Association of Universities for Research in Astronomy (AURA) under a cooperative agreement with the National Science Foundation. This Letter makes use of the ALMA data from ADS/JAO.ALMA\#2017.1.00001.S. ALMA is a partnership of ESO (representing its member states), NSF (USA) and NINS (Japan), together with NRC (Canada) and NSC and ASIAA (Taiwan)  and KASI (Republic  of Korea), in cooperation with the Republic of Chile. The Joint ALMA Observatory is operated by ESO, AUI/NRAO and NAOJ. W.R. acknowledges support from Chulalongkorn University's CUniverse and the Ratchadapiseksompot Endowment Fund.

\end{CJK*}

\begin{thebibliography}{}
\bibitem[Barro et al.(2016)]{Barro16} Barro, G., Kriek, M., P{\'e}rez-Gonz{\'a}lez, P.~G., et al.\ 2016, \apjl, 827, L32 
\bibitem[Chen et al.(2022)]{Chen22} Chen, C.-C., Gao, Z.-K., Hsu, Q.-N., et al.\ 2022, \apjl, 939, L7. doi:10.3847/2041-8213/ac98c6
\bibitem[Cheng et al.(2023)]{Cheng23} Cheng, C., Huang, J.-S., Smail, I., et al.\ 2023, \apjl, 942, L19. doi:10.3847/2041-8213/aca9d0
\bibitem[Cibinel et al.(2017)]{Cibinel17} Cibinel, A., Daddi, E., Bournaud, F., et al.\ 2017, \mnras, 469, 4683
\bibitem[Dunlop et al.(2017)]{Dunlop17} Dunlop, J.~S., McLure, R.~J., Biggs, A.~D., et al.\ 2017, \mnras, 466, 861 
\bibitem[Elbaz et al.(2018)]{Elbaz18} Elbaz, D., Leiton, R., Nagar, N., et al.\ 2018, \aap, 616, A110 
\bibitem[Faure et al.(2021)]{Faure21} Faure, B., Bournaud, F., Fensch, J., et al.\ 2021, \mnras, 502, 4641
\bibitem[Fensch \& Bournaud(2021)]{Fensch21} Fensch, J. \& Bournaud, F.\ 2021, \mnras, 505, 3579
\bibitem[Fujimoto et al.(2018)]{Fujimoto18} Fujimoto, S., Ouchi, M., Kohno, K., et al.\ 2018, \apj, 861, 7
\bibitem[Hodge et al.(2019)]{Hodge19} Hodge, J.~A., Smail, I., Walter, F., et al.\ 2019, \apj, 876, 130
\bibitem[Hogg \& Lang(2013)]{HoggLang13} Hogg, D.~W., \& Lang, D.\ 2013, \pasp, 125, 719
\bibitem[Iono et al.(2016)]{Iono16} Iono, D., Yun, M.~S., Aretxaga, I., et al.\ 2016, \apjl, 829, L10 
\bibitem[Ivison et al.(2020)]{Ivison20} Ivison, R.~J., Richard, J., Biggs, A.~D., et al.\ 2020, \mnras, 495, L1
\bibitem[Kalita et al.(2022)]{Kalita22} Kalita, B.~S., Daddi, E., Bournaud, F., et al.\ 2022, \aap, 666, A44
\bibitem[Kaasinen et al.(2020)]{Kaasinen20} Kaasinen, M., Walter, F., Novak, M., et al.\ 2020, \apj, 899, 37
\bibitem[Nelson et al.(2019)]{Nelson19} Nelson, E.~J., Tadaki, K.-i., Tacconi, L.~J., et al.\ 2019, \apj, 870, 130 
\bibitem[Mandelker et al.(2017)]{Mandelker17} Mandelker, N., Dekel, A., Ceverino, D., et al.\ 2017, \mnras, 464, 635 
\bibitem[Pichardo et al.(2003)]{Pichardo03} Pichardo, B., Martos, M., Moreno, E., et al.\ 2003, \apj, 582, 230
\bibitem[Puschnig et al.(2023)]{Puschnig23} Puschnig, J., Hayes, M., Agertz, O., et al.\ 2023, arXiv:2303.13858. doi:10.48550/arXiv.2303.13858
\bibitem[Rizzo et al.(2020)]{Rizzo20} Rizzo, F., Vegetti, S., Powell, D., et al.\ 2020, \nat, 584, 201
\bibitem[Rodighiero et al.(2011)]{Rodighiero11} Rodighiero, G., Daddi, E., Baronchelli, I., et al.\ 2011, \apjl, 739, L40 
\bibitem[Rujopakarn et al.(2016)]{Rujopakarn16} Rujopakarn, W., Dunlop, J.~S., Rieke, G.~H., et al.\ 2016, \apj, 833, 12
\bibitem[Rujopakarn et al.(2019)]{Rujopakarn19} Rujopakarn, W., Daddi, E., Rieke, G.~H., et al.\ 2019, \apj, 882, 107
\bibitem[Schreiber et al.(2015)]{Schreiber15} Schreiber, C., Pannella, M., Elbaz, D., et al.\ 2015, \aap, 575, A74 
\bibitem[Simons et al.(2019)]{Simons19} Simons, R.~C., Kassin, S.~A., Snyder, G.~F., et al.\ 2019, \apj, 874, 59. doi:10.3847/1538-4357/ab07c9
\bibitem[Speagle et al.(2014)]{Speagle14} Speagle, J.~S., Steinhardt, C.~L., Capak, P.~L., \& Silverman, J.~D.\ 2014, \apjs, 214, 15 
\bibitem[Spergel(2010)]{Spergel10} Spergel, D.~N.\ 2010, \apjs, 191, 58
\bibitem[Swinbank et al.(2015)]{Swinbank15} Swinbank, A.~M., Dye, S., Nightingale, J.~W., et al.\ 2015, \apjl, 806, L17 
\bibitem[Tadaki et al.(2017)]{Tadaki17} Tadaki, K.-. ichi ., Genzel, R., Kodama, T., et al.\ 2017, \apj, 834, 135
\bibitem[Tadaki et al.(2018)]{Tadaki18} Tadaki, K., Iono, D., Yun, M.~S., et al.\ 2018, \nat, 560, 613 
\bibitem[Tamburello et al.(2017)]{Tamburello17} Tamburello, V., Rahmati, A., Mayer, L., et al.\ 2017, \mnras, 468, 4792 
\bibitem[Whitaker et al.(2012)]{Whitaker12} Whitaker, K.~E., van Dokkum, P.~G., Brammer, G., \& Franx, M.\ 2012, \apjl, 754, L29 
\bibitem[Williams et al.(2023)]{Williams23} Williams, C.~C., Tacchella, S., Maseda, M.~V., et al.\ 2023, arXiv:2301.09780. doi:10.48550/arXiv.2301.09780
\bibitem[Wu et al.(2023)]{Wu23} Wu, Y., Cai, Z., Sun, F., et al.\ 2023, \apjl, 942, L1. doi:10.3847/2041-8213/aca652
\end{thebibliography}
\end{document}